\documentclass[prl,twocolumn,amssymb]{revtex4}

\begin{document}

\title{Reply to Comment on ``Electron Mass Operator in a Strong Magnetic Field 
and Dynamical Chiral Symmetry Breaking''}

\author{A. V. Kuznetsov}
 \email{avkuzn@uniyar.ac.ru}
\author{N. V. Mikheev}
 \email{mikheev@uniyar.ac.ru}
\affiliation{
Division of Theoretical Physics, Department of Physics,\\
Yaroslavl State University, Sovietskaya 14,\\
150000 Yaroslavl, Russian Federation}

\date{July 10, 2002}

\maketitle

In the Letter~\cite{Kuznetsov:2002prl} we have demonstrated 
a consistent calculation of the electron mass operator in a strong 
magnetic field in the leading-log approximation (LLA). 
It means that we have always neglected the terms 
$\sim \alpha^2 \ln (e B/m^2)$ and $\sim \alpha$ against the terms
$\sim \alpha \ln (e B/m^2)$. This program has been successfully 
realized with the main result presented in Eq.~(11). 

The analysis has shown that the solution of Eq.~(11) for the 
electron physical mass $m$ was independent in strong fields 
on the initial field-free mass $m_0$ and was reduced in fact 
to the solution at $m_0 = 0$. 
The result (11) has been applied to a calculation of the fermion
dynamical mass generated by a magnetic field, in a model 
with $N$ charged fermions which were massless in the field-free limit.
This treatment has led us to such unusual conclusions as an existence 
of a critical number of fermions in the theory, $N_{cr}$, 
such that the dynamical mass has not been arised for $N > N_{cr}$ 
and it has been generated with a doublet splitting for $N < N_{cr}$. 

The authors of the Comment~\cite{Gusynin:2002} show, that taking 
into consideration of {\it a part} of the next-to-leading corrections 
changes the results fundamentally: no $N_{cr}$ and no double splitting 
arise. 

However, one can see that using of the ``corrected'' formula~(2) of 
Ref.~\cite{Gusynin:2002} instead of our approximate formula (12)
of Ref.~\cite{Kuznetsov:2002prl} is a little part of a step beyond 
the LLA. Really, the left-hand side of Eq.~(2), Ref.~\cite{Gusynin:2002}, 
can be rewritten as
\begin{eqnarray}
\left[\frac{\alpha_R}{2 \pi} 
\left(\ln \frac{\pi}{N \alpha_R} - \gamma_{\text{E}} \right) 
+ O (\alpha_R^2) \right] \,
\ln \frac{e B}{m^2}.
\label{eq:expan}
\end{eqnarray}
On the other hand, such the next-to-leading terms 
$\sim \alpha_R^2 \ln (e B/m^2)$ 
have already been neglected, for example, when the truncated set 
of the Schwinger-Dyson equations was formulated. Namely, 
the reduction of the exact vertex in a strong magnetic field to the bare 
one, $\Gamma_\mu \to \gamma_\mu$, is valid in LLA only.

To give a conclusion on the dynamical mass generation by a strong 
magnetic field in the next-to-leading log approximation, one should 
consistently take such $\alpha^2 \ln$ corrections everywhere, and first, 
without the truncation of the Schwinger-Dyson equation set. 
To our knowledge, this program was not performed yet.

Regarding the renormalization of the coupling constant $\alpha_R$, 
it follows directly (see, e.g.~\cite{Kuznetsov:2002mpl})
from the photon polarization operator in a strong 
magnetic field:
\begin{eqnarray}
{\cal P} = \frac{N \alpha}{3 \pi}\;q^2\;f (\dots)
- \frac{2 N \alpha}{\pi} \; eB \exp \!\!\left(\!-\frac{q_{\bot}^2}{2 e B}\right)
H \!\!\left(\!\frac{q_{\|}^2}{4 m^2}\right)\!,
\label{eq:P}
\end{eqnarray}
where the function $H(z)$ is presented in Eq.(9) of our Letter,
and the function $f (q_{\|}^2,q_{\bot}^2,m^2,e B)$ has in general case
the form of a double integral over the Fock-Schwinger proper times,
and can be extracted from the papers by Tsai and Shabad 
cited in our Letter. In the limiting case regarding to our consideration, 
$q_{\|}^2,m^2 \ll q_{\bot}^2 \ll e B$, the function $f$ is simplified, 
$f \simeq \ln (e B/m^2) - 1.792$. It immediately leads in the LLA to 
Eq.(13) of our Ref.~\cite{Kuznetsov:2002prl}. 
In another limit, e.g. $q_{\bot}^2 \gg e B$, one has 
$f \simeq \ln (q_{\bot}^2/m^2) - 5/6$. 
In the LLA $\alpha_R$ appears to look like 
a field-free running coupling with the substitution $\mu^2 \to e B$, 
but in a general case it is a complicated function 
of the field strength and of the momenta $q_{\bot}^2,q_{\|}^2$. 
Thus, the statement made in the Comment that 
we took the running coupling from the one-loop RG equations in QED 
without a magnetic field, is incorrect. 

We note also, that the specificity of our approach is the 
self-regulation of the 
``condition'' $\alpha_R < \infty$ emphasized in the Comment.

We did not speculate in our Letter on possible physical 
consequences of an appearance of two different dynamical masses. 
Possibly, it would lead not only to two different theories, 
but rather to two different types of domains of our Universe.

\bibliography{repl_com}

\end{document}